\documentclass[12pt,preprint]{aastex}
\usepackage{rotating}
\usepackage{changepage}

\newcommand{\ltsim}{\raisebox{-1.0ex}{$\stackrel{\textstyle<}{\sim}$}}
\newcommand*\diff{\mathop{}\!\mathrm{d}}
\def\kms{km~s$^{-1}$}

\def\goes{{\sl GOES}}
\def\yohkoh{{\sl Yohkoh}}

\def\hinode{{\sl Hinode}}

\def\sdo{{\sl SDO}}

\def\p78{{\sl P78-1}}

\def\iris{{\sl IRIS}}

\def\cii{C~{\sc ii}}
\def\siiv{Si~{\sc iv}}
\def\mgii{Mg~{\sc ii}}
\def\halpha{H$\alpha$}

\def\kms{km~s$^{-1}$}

\def\etal{et~al.}

\begin{document}
%
\slugcomment{Accepted for publication in \underline{The Astrophysical Journal}}

\title{Solar Active Region Coronal Jets II:  Triggering and Evolution of Violent Jets}

\author{Alphonse C.~Sterling\altaffilmark{1}, Ronald L. Moore\altaffilmark{1,2}, David A. Falconer\altaffilmark{1,2}, Navdeep K. Panesar\altaffilmark{1}, and 
Francisco Martinez\altaffilmark{3}}

\altaffiltext{1}{NASA Marshall Space Flight Center, Huntsville, AL 35812, USA,\newline alphonse.sterling@nasa.gov, ron.moore@nasa.gov}
\altaffiltext{2}{Center for Space Plasma and Aeronomic Research, University of Alabama in Huntsville, AL 35899, USA}
\altaffiltext{3}{Georgia State University, Atlanta, GA, USA}


\begin{abstract}

We study a series of X-ray-bright, rapidly-evolving active-region coronal jets outside the leading sunspot
of AR~12259, using \hinode/XRT, \sdo/AIA
and HMI, and \iris\ data.  The detailed evolution of such  rapidly evolving ``violent'' jets  remained a
mystery after our previous investigation of active region jets   \citep[][Paper~1]{sterling_et16}. The 
jets we investigate here erupt from three localized subregions, each
containing a rapidly evolving (positive) minority-polarity
magnetic-flux patch bathed in a (majority) negative-polarity magnetic-flux background. At least several of
the jets begin with eruptions of what appear to be thin (thickness $\ltsim 2''$)  miniature-filament
(minifilament) ``strands'' from a magnetic neutral line where magnetic flux cancelation is ongoing, consistent 
with the
magnetic configuration presented for coronal-hole jets in  \citet[][]{sterling_et15}.  
Some jets strands are difficult/impossible to detect, perhaps due to, e.g.\ their thinness, obscuration by
surrounding bright or dark features, or the absence of erupting cool-material minifilaments in those jets. 
Tracing in detail the flux evolution in one of the subregions, we find bursts of strong jetting occurring only
during times of  strong flux cancelation.  Averaged over seven jetting episodes, the cancelation rate was
$\sim$1.5$ \times 10^{19}$~Mx~hr$^{-1}$.  An average flux of $\sim$5$\times 10^{18}$~Mx canceled prior to each
episode, arguably building up $\sim$10$^{28}$---10$^{29}$~ergs of free magnetic energy per jet. From these and previous
observations, we infer that flux cancelation
is the fundamental process responsible for the pre-eruption buildup and triggering of at least many jets in
active regions, quiet regions, and coronal holes.


\end{abstract}


\section{Introduction}
\label{sec-introduction}

Solar coronal jets are dynamic features that grow to long lengths compared to their widths.  They occur in all solar
environments: quiet Sun, coronal holes, and the edges of active regions \citep{raouafi_et16}.   \yohkoh's soft X-ray
telescope (SXT) primarily saw them in active regions \citep[e.g.][]{shibata_et92,shimojo_et96}, while \hinode's X-ray
telescope (XRT) found them to be plentiful in polar coronal holes also \citep{cirtain_et07}.  In polar coronal holes
they typically reach heights of $\sim$50{,}000~km and widths of $\sim$8000~km, occurring at a rate of $\sim$60/day in
the two polar holes, and having lifetimes of $\sim$10~min \citep{savcheva_et07}.  \citet{pucci_et13} found such jets
to have energies  $\sim$10$^{26}$---10$^{27}$~erg for the hotter component of the jet, and they point out that the
chromospheric/cool-coronal jet that sometimes accompanies the hot component \citep{moore_et10,moore_et13} could 
increase this by about an order of magnitude. In a study weighted toward active region jets (where 68\% of 100 jets 
occurred in active regions, and the others in either coronal holes or quiet Sun),  \citet{shimojo_et96} report: a
large range of lengths, averaging $1.5\times10^5$~km; a large range of lifetimes, going from $\sim$100~s to $\sim
4$~hrs; and outward velocities ranging over 100---400~\kms, with an average of 200~\kms. For a sample of jets,
\citet{shimojo_et00} found energies of $\sim$10$^{27}$---$\sim$10$^{29}$~erg.  Independent of where they occur on the
Sun, essentially all coronal jets observed in X-rays appear as a bright spire with a base brightening, often with the
brightest part of the base sitting on one side of the base; we call this a jet-base bright point (JBP)\@.   (In EUV
images the JBP is also usually apparent as a sharp brightening, although at times it does not stand out as  distinctly
as it does in X-rays, due to factors such as the details of the EUV passbands and the possible presence of obscuring
cooler-temperature material.)

Explaining the JBP in addition to the spire has been a key aspect of proposed physical mechanisms for what
produces the jets.  An early suggestion was that newly emerging magnetic flux elements reconnecting with
ambient coronal fields produces the jets \citep[e.g.,][]{shibata_et92,yokoyama_et95}.  This emerging-flux
model for coronal jets was put forth when the best data available in which to see jets were X-ray images,
and the cadence and regularity of magnetic data was limited.  More recent data sources, such as the
multiple-EUV-wavelength images from the Atmospheric Imaging Assembly (AIA) and the regular and
high-cadence magnetograms from the Helioseismic and Magnetic Imager (HMI), both on the Solar Dynamics
Observatory (\sdo) satellite, have made it possible to examine the origins of coronal jets in greater
detail.  Images from AIA show that, at least in many cases, the jets originate from the eruption of a
small-scale filament \citep[e.g.][]{shen_et12,liu_et14,adams_et14,sterling_et15}; following earlier works,
we will call these small-scale filaments ``minifilaments.''  Figure~1 shows a schematic of the
minifilament-eruption jet-production process as proposed by \citet{sterling_et15}.

This minifilament-eruption picture however was originally developed from observations of polar coronal
hole jets \citep{sterling_et15}.  We have recently found the picture also to hold for quiet Sun coronal
jets \citep{panesar_et16a}.  \citet{sterling_et16} (Paper~1 hereafter) investigated whether {\it active
region jets} also typically follow the Figure~1 scenario, and the findings were mixed.  They observed
several jets, and found that at least three of them showed characteristics of the minifilament-eruption
scenario.  These three jets were studied in detail in Paper~1; they showed relatively slow development,
with a cool-material minifilament well-resolved by AIA in time and space as it moved toward eruption and
jet formation.  Two of  these events made a surge and a spray, but showed either a  relatively weak
signature or no obvious signature in available X-ray images (obtained from the Soft X-ray Imager, SXI;
\hinode\ XRT data not being available).   The third slowly evolving jet did produce a stronger X-ray
signature, although still modest compared to other jets.  All three of those three jets could be
understood with the minifilament-eruption interpretation of Figure~1.    In addition however, there were
several jets that made strong SXI X-ray jets (originating from region ``C'' of Fig.~3(a) in Paper~1).  
Paper~1 did not determine the origin of those jets because they were fast-developing and/or because their
early-development stages were obscured by surrounding material.  In this paper we refer to such
fast-developing jets as ``violent'' active region jets, to distinguish them from the
easier-to-decipher more-gradually-developing jets that clearly resulted from erupting minifilaments. 
Paper~1 was however able to conclude that those violent jets did occur at a neutral line that showed
unambiguous flux cancelation. 

Thus, from the Paper~1 study, the question of whether those comparatively violent jets
follow the minifilament-eruption scheme was unresolved.  Here we revisit this question by studying
similarly violent jets from a different active region.  We use a broader range of data than in the
previous study, including images from the \hinode\ X-ray telescope (XRT) and from the Interface Region
Imaging Spectrograph (\iris) instrument, as well as \sdo/AIA and HMI, to assist with trying to unravel
the nature of these violent active region coronal jets.

\section{Instrumentation and Data}
\label{sec-instruments}

For our investigations we use \sdo/AIA \citep{lemen_et12} images in seven EUV wavelength bands: 304~\AA,
171~\AA, 193~\AA, 211~\AA, 131~\AA, 94~\AA, 335~\AA, and in one UV band: 1600~\AA\@.  Each of these
channels tends to detect plasma emissions within a limited temperature range determined by characteristic
emission  lines around these wavelengths, although in some cases there can be two or more such
characteristic temperatures; see \citet{lemen_et12} for details.  AIA observes the full Sun, has detectors
with $0''.6$ pixels, and nominally operates with a cadence of 12~s in the EUV channels and 24~s in the UV
channels.  \hinode/XRT \citep{golub_et07} has a broadband response in the X-ray range to emissions from
plasma hotter than about $1.0 \times 10^6$~K\@.  It has $1''.02$ pixels, and normally observes the Sun 
with a limited field of view (FOV) and with varying cadences. \iris\ \citep{depontieu_et14} does high
resolution ($0''.17$ pixels) imaging in the UV range with its slit-jaw camera, with four different
filters: 1330~\AA\ (\cii, 30{,}000~K) and 1400~\AA\ (\siiv, 65{,}000~K) with a 40~\AA\ bandpass each, and
2796~\AA\ (\mgii~k, 15{,}000~K) and 2831~\AA\ (\mgii~h/k wing, 6{,}000~K) with a 4~\AA\ bandpass each. For
the time period we investigate in Section~\ref{sec-r2}\ below, slit-jaw images are available in  all but
the 2831~\AA\ line.

\section{Data Set: Violent Jets in X-rays and in EUV}
\label{sec-data}

Our data selection is based on XRT observations from the second-half of the disk passage of NOAA
AR~12259, from 2015 January 13 12:34~UT to January 20 06:30~UT\@.  This is during a period of ``focused mode''
observations by \hinode, when relatively-long-term observations on selected targets are carried out. 
Although the XRT observations were not completely continuous over this period, there were long stretches
of uninterrupted observations, typically at a cadence of $\sim$5~min, although with some periods of
higher cadence (30~s). Figure~2 shows the region in white
light, along with a magnetogram.  Over this period there were numerous X-ray jets occurring at the
periphery of the region, with the jets tending to cluster in different locations of the region for
extended periods of time.  In this paper, we will focus on one such cluster of jets, occurring in  the
boxed region of Figure~2(a) over 2015 January 13-14.  

Figure~3 shows sample XRT images of the region, and the accompanying animation shows the region over the
full period of our observations.  XRT had a gap in observations during the period, with the first set of
observations running between 13 January 14:34~UT and 17:44~UT, and the second  running nearly
uninterrupted between 14 January 06:16~UT and 16 January 09:43~UT\@.  For our investigations, we cut off
this second block at 14 January 22:34~UT,  because the jetting in the Figure~3 FOV largely ends around
that time. From the animation accompanying Figure~3, one can see that the jets happened in the three
subregions, R1, R2, and R3 of Figure~3.  

For the study of this paper, we examined jets occurring in the FOV of Figure~3 over the time period 
from 2016 January 13 14:30~UT to January 14 23:35~UT\@.  We discuss in most detail (see Sect.~\ref{sec-r2} onward) 
jets of the R2 region for a more restricted period, on 2016 January 14 from 12:35~UT 
to 22:35~UT, when that region was particularly jet productive.

Figure~4 shows the three jetting regions in AIA EUV 193~\AA\ and AIA~94~\AA\ images, with and without overlaid
HMI magnetograms.  We also
include accompanying  two-minute-cadence movies that, different from the XRT movies of Figure~3, are
uninterrupted over the entire time period of our observations.  From these movies we can see that jets occur
nearly continuously throughout the period, although there still is a quieter period on Jan~14 between
about 16:00~UT and 17:20~UT, as is also apparent in the XRT movie.  

Figure~5 shows an intensity lightcurve from the AIA 94~\AA\ channel over the observation period, where we have integrated
the intensity at each timestep over the FOV of the Figure~3 and Figure~4 images; thus this  lightcurve shows the
intensity variations over R1, R2, and R3 regions of Figure~3.  If XRT had the cadence of AIA, some of the jets seen by AIA
would undoubtedly have been detected by XRT in-between  the times of some of those visible in the XRT animation
accompanying Figure~3.

Unlike the active region jets of Paper~1, most jets of this study were not clearly identifiable in the
\goes\ soft X-ray flux plots. This is largely due to a somewhat higher \goes\ background for these events;
for the times of our events here the background is very near the C1 level, while for the events of Paper~1
the average background was at the high-B level.  Additionally, at least one other region, AR~12257 at the
west limb, was very active during this period, which might contaminate \goes\ emissions near the time of
our identified jets.  Two events identified by NOAA as flares that are likely in the base of our jets are
a C1.6 and a C1.9 event at 01:54~UT and 10:38~UT, respectively, on 2015 January~14.

It is apparent from the 193~\AA\ animation accompanying Figure~4(a) that fan loops that emanate from the
sunspot on the south edge of the figure obscure much of the jet activity, making it difficult to see the
locations of origin of many of the jets. Also, absorbing material, frequently retracting jet material,
sometimes obscures observations of the jets; Paper~1 also identified this as a difficulty with trying to
observe the origin of some of  their jets.

Overlaying HMI magnetograms onto the images (Figs.~4(c) and 4(d)) shows that there is a mixture  of
polarities in the region.  Overall, the dominant polarity is negative, but there are three islands of 
positive polarity, the minority polarity of the region.  These three mixed-polarity regions coincide
with the three X-ray-bright areas of Figure~3: R1, R2, and R3, and are the source of the bulk of the
jets of the region over the time period covered by the movies accompanying Figures~3 and~4.  In the
following we will first consider a set of jets occurring in region R2, and then we will consider
examples of jets from  the other locations.

\section{Homologous Violent Jets from Subregion R2}
\label{sec-r2}

Subregion~R2 of Figure~3 shows particularly substantial jetting activity on January~14, with many of the
jets apparently homologous.  We first  examine in detail a jet episode of January~14 that begins in XRT
images near 14:40~UT\@.  We select this particular event because we had excellent coverage of it; in addition 
to AIA and HMI data, we also have available \iris\ slit-jaw images and high-cadence
($\sim$30 s) XRT images.

Figure~6 shows the jet region.  Over the period of the high-cadence AIA movies accompanying Figure~6(a),
there is actually a sequence of several repeated jets around the time of the strong one near 14:40~UT\@.
From the animation accompanying Figure~6(a), each jet begins at an isolated location that becomes the
eastern edge of the jet, near (x,y)=(85,-150).  A very weak jet starts at 13:50~UT and continues for about
7 min.  Another starts from the same location at $\sim$14:07:11~UT, with the emission from that developing
jet remaining confined to a location with  narrow extent in the  east-west direction.  It then however
spreads toward the west with time, first gradually and then relatively explosively from 14:17~UT, and
reaches a maximum  extent of $\sim$15$''$ at about 14:30~UT.  As this jet episode is fading, the
above-mentioned jet that starts at 14:40~UT in X-rays   begins; in the AIA videos, its initial brightening
starts  at about 14:32~UT at the eastern location.  It starts to broaden markedly from about 14:41~UT,
reaches a maximum extent near 14:48~UT, and then fades.  A new feature then brightens from 14:52~UT, but
it is from a  different origin site, and so is not homologous with the three events just described.  But a
new event, homologous with those three earlier events, does start at about 15:02~UT, although it is
heavily obscured by foreground opaque material; it has a more complicated development that is still
continuing as the Figure~6(a) high-cadence movie ends at 15:40~UT\@.

Figure~6(b) shows the magnetogram of the region, and Figure~6(c) shows an \iris~1400~\AA\ image of part 
of the FOV of the images in the earlier two panels.  Figures~6(d)---6(f) show the \iris\ image at
three different times with the magnetogram contours overlaid.  This sequence begins in Figure~6(d) with
the above-mentioned event that started in AIA at 14:32~UT\@.  It is clear
that the eastern edge of the event is lodged into the neutral line between polarities B3 and B4 in
Figure~6(d), with  negative/positive polarities on the east/west side of the originating jet. On its
southern edge the jet may again be bounded by a less-well-defined neutral line that includes the positive
flux clump labeled B6 in Figure~6(d).  Figure~6(f) shows the jet at near its maximum extent; at this time
its western edge is lodged in a region of negative polarity.

We can make a direct comparison between the \iris-observed jet and magnetograms in Figures~6(d)---6(f), and the
jet schematic of Figure~1.  In both the schematic and in the observations, the dominant polarity is
negative and the minority polarity is positive.  In the schematic, the jet starts between the respective 
negative and  positive polarities ``A'' and ``B'' of Figure~1(c), and then in time the jet spreads
horizontally out to polarity ``C\@.''  Similarly, in the observations, the  jets starts between B3 and B4
(Fig.~6(d)), and spreads out to the location of B5 (Fig.~6(f)).  Therefore, the observations of this
active region jet are broadly consistent with the schematic of Figure~1, which was drawn based on
observations of coronal hole jets.   We also inspected other available \iris\ slit-jaw wavelengths for
this event, and they are consistent with what we find from the 1400~\AA\ images.

One question about this comparison between these violent active region jet observations and the Figure~1
schematic however is whether there is a minifilament that erupts from the Figure~6(d) B3---B4 neutral 
line to form the
jet, as presented in the Figure~1 schematic.  In AIA EUV images, for the jet episode beginning at about
14:07~UT,   we can identify strands of absorbing material erupting from the B3---B4 neutral line.  One
such  strand is touched by the tip of the black arrow  in the 171~\AA\ image in Figure~6(a); it has width
$\ltsim 2''.$  A similar such strand is discernible in a subsequent jetting episode in the video
accompanying Figure~6(a) at times around 14:41:23~UT\@. When viewed against the disk however, filament 
material is apparently difficult to see in \iris\ slit-jaw images, reportedly being semi-transparent and
optically thin in 1330~\AA\ and 1400~\AA\ images \citep{li_et16}.  Consistent with this, we  see no
absorbing filament material at those slit-jaw wavelengths, nor do we see any such material in  the
2796~\AA\ slit-jaw images.  Nonetheless, the AIA 171~\AA\ (and also 193~\AA) observations are suggestive 
of ejections of cool
minifilament material in narrow, long strands.  While these features are only marginally apparent here, we
will show cases below (Section~\ref{sec-minifilaments}) where thin filament-like strands appear more
clearly than here.

Also to be consistent with the Figure~1 schematic, a JBP should occur along the B3---B4 neutral line.
We identify candidate brightenings in AIA 94~\AA\ images  (video accompanying Fig.~4(b)), e.g.\ at
14:16:01~UT and again at about 14:40:01~UT and about 15:16:01~UT in  the video accompanying
Figures~4(b) and~4(d).  This brightening is also apparent in X-rays, e.g.\ on Jan~14 at 14:20:52~UT, 14:44:32~UT,
and at 15:19~UT in the video accompanying Figure~3; the location of this  brightening matches that of
the JBP in the Figure~1(b)---1(c) schematic.  Our observed candidate brightenings however are soon
superseded in brightness by neighboring loops rooted in stronger magnetic fields. 

Similarly, we expect new bright loops connecting the minifilament-eruption location, i.e.\ the
minority-polarity flux near the B3---B4 neutral line, with the majority-polarity roots of the neighboring
larger loop in the Figure~1 schematic, at the times of Figures~1(c) and~1(d).  Such connections are
repeatedly identifiable in the AIA 94~\AA\ video accompanying  Figures~4(b) and~4(d) at, e.g., on Jan~14 
around 14:26:01~UT, 14:46:01~UT, and at 15:26:01~UT, respectively with the brightenings listed in the previous
paragraph.  These connections also  are apparent in the \iris\ image in Figure~6(f), where the jet
intensity brightenings clearly extend from the positive-polarity side of the B3---B4 neutral line to the
B5 negative-polarity region.  In each of these cases, the lateral extent of bright emitting jet material
expands to cover the approximate extent of the base region, consistent with the shadings in Figure~1(c)
and~1(d).  

This spire expansion with time is also apparent in the XRT X-ray video accompanying Figure~3,
especially for the higher-cadence observations of the jet over 14:41~UT---14:51~UT, and thus conforms
to the definition of blowout jets presented in \citet{moore_et10,moore_et13}.  Moreover, a prediction
of the minifilament-eruption scenario is that the jet spire should move away from the JBP location with
time \citep{sterling_et15}; this direction of movement is clearly visible in the XRT movie (and other
movies) for this jet.

We conclude that for the time interval of the \iris\ movie of Figure~6 (approximately
14:48~UT---15:25~UT on Jan~14), in most respects the jets originating at the B3---B4 neutral line (labeled in
Fig.~6(d))
follow the schematic  of Figure~1.  Possible differences however are that the  emission at the expected
JBP is soon dominated in brightness by emissions from neighboring stronger-field loops.  Also, we are
not able to determine unambiguously whether there is a cool-material minifilament ejected at the start
of the eruption.  We do observe narrow strand-like absorbing features near the AIA spatial-resolution
limit, that are embedded/entrained in bright jetting material, and could be an erupting minifilament. 
We next consider other jets in the same active region where such  strands unambiguously
participate in the jet development in the manner depicted in Figure~1.

\section{Violent Jet Eruptions Displaying Obvious Erupting Minifilaments}
\label{sec-minifilaments}

In Section~\ref{sec-r2} we focused on events for which we had the best mutual coverage among XRT and
\iris, in addition to AIA and HMI\@.  Those jets however did not show clearly an accompanying minifilament
eruption.  Many other jets however occur among the three bright areas of this active region depicted in
Figure~3.  From the low-cadence AIA 193~\AA\ video accompanying Figure~4(a), we can identify several cases
where we can discern minifilaments erupting in step with the jet more clearly than in the examples of
Section~\ref{sec-r2}. 

Figure~7 shows two such jets; although XRT and \iris\ data are not available at these times, we present
AIA 193~\AA\ images in Figure~7 and 12~s-cadence images in the accompanying animations.  There are two
successive jet eruptions, respectively starting at $\sim$01:34~UT and at $\sim$01:50~UT\@.   in this 
case the jets originate from
subregion R1, and specifically from the B1---B2 neutral line and spread out to the B3 polarity
(Fig.~7(d); see Fig.~6(d) for labels).  This is in contrast to the situation in Figure~6, where the jets 
originated in subregion R2, and specifically from the
B3---B4 neutral line, and spread out to the B5 polarity.  That is, for the events of both Figures~6 and~7, the 
jets form with a positive-flux
minority-polarity setup.  In the Figure~6 case that positive polarity is B4,  and in the Figure~7 case that
positive polarity is B2.  Thus  in Figure~7, in comparing with the labeling of the Figure~1(c) schematic,
locations B1, B2, and B3 correspond to locations ``A,'' ``B,'' and ``C'' of the schematic, respectively.

Polarities of these subregions however are rapidly evolving.  So at the time of Figure~7 the
positive-polarity patch labeled B4 in Figure~6(d) has not yet formed, and by the time of Figure~6 the
negative-polarity patch labeled B1 in Figure~7(d) is no longer present.  We will discuss these rapid evolutionary
magnetic changes in Section~\ref{sec-b evolution}\ below.

In contrast to the events of Section~\ref{sec-r2}, here {\it we can discern clearly} an erupting minifilament  in the
magnetic eruption that drives each jet.  From the Figure~7(a) video, this minifilament becomes discernible from the
background dark material from about 01:29:06~UT\@.  It then ``crawls up'' the neighboring mound-like bright region,
corresponding to the time period between the schematics in Figures~1(a) and~1(b), until about 01:31:42~UT\@.  It then
starts moving (approximately)  radially outward from the surface along what is to become the jet spire; this upturn in the
minifilament's trajectory occurs at a time corresponding to that between the schematics in Figures~1(c) and~1(d).  In the
animation accompanying Figure~7(a), dark material from the outward-moving minifilament quickly becomes engulfed in
surrounding hotter jet material; this hotter material,  represented by the shaded regions in Figures~1(c) and~1(d), is
expected to form when field enveloping the field threading the cool minifilament material undergoes external reconnection
with far-reaching field of the majority polarity, which is negative in both the observations and in the schematic.  As with
the jet of Figure~6, in this case the jet spire also starts out relatively narrow (e.g., at 01:35~UT), but then expands to
span much of the mound-like bright region (e.g., at 01:37~UT), corresponding to the situation of the schematic in
Figure~1(d).  Again, in accordance with that schematic, the expansion is  in a direction away from where the minifilament
originally resides prior to its eruption.

For the second jet in the movie of Figure~7, the filament is even more obvious, first appearing at about
01:45:06~UT, and starting to erupt upward from about 01:50:18~UT\@.  

These minifilaments appear to be strands of width $\ltsim 2''$.  This is the same size range we inferred
for jets of Figure~6 also, but they are much more obvious (darker) for these Figure~7 events.  This could
possibly be due to a better perspective in this case; here we are viewing region R1 instead of R2,
allowing us to see more of the minifilaments as they crawl up the neighboring mound-like regions.  Another
factor could be that in the case of Figure~7 we can see the minifilaments before much of the external
reconnection occurs between the field enveloping the minifilament and the surrounding coronal field; the
hot emissions from that reconnection tend to create very bright jet-spire and large-lobe-loop material
that blinds our view of the cooler material.

We have also inspected 12~s-cadence AIA 94~\AA\ images of this same event in order to confirm the
location of the JBP, but the findings regarding this are equally apparent in the 193~\AA\ images of
Figure~7 and its accompanying video, and thus we do not include the 94~\AA\ images for this case. 
Based on the schematic of Figure~1, we expect the JBP to occur at the B1---B2 neutral line.  For the
first of the two jetting episodes, there is only very weak brightening in 193~\AA\ at this location at
about 01:20:54~UT\@. For the second jet episode, brightening at this neutral line (e.g., over
approximately 01:48~UT---01:52~UT) is a little stronger than in the first episode.  

In both episodes however, the strongest base brightening occurs $\sim$5$''$---10$''$ further to the
west of the B1---B2 neutral line from which the minifilament strand originates.  For the first jet
episode this strongest brightening is  at about 01:37:30~UT\@.  For the second jet episode, at the time
of the strongest brightenings the images are too saturated to make a reliable match to the magnetogram,
but at, e.g., 02:18~UT, it is clear that the main base brightening is west of the B1---B2 neutral
line.  In both cases, these brightenings are strongest at about the time the minifilaments lift off
radially away from the surface.  For these jets, B2 is a strong positive polarity, and it is surrounded
by other negative elements in addition to B1, including the negative-polarity patch B3, and negative
patches north, south, and southeast of B3.  These brightening locations are consistent with them being
intense newly-formed loops from the external reconnection of the envelope of an erupting minifilament flux rope,
corresponding to the red loop connecting polarities B and C in Figure~1(c); they are particularly
bright because the flux elements in which those loops are rooted are particularly strong. 

Another possibility is that the strong brightenings west of the B1---B2 neutral line is actually a 
secondary-jet-causing eruption, triggered by the filament eruption from the B1---B2 neutral line.  In
that case, that brightening would form the JBP for another jet, with the neutral line of that JBP
being between the positive patch B2 in Figure~7(d) and negative polarity south of B2 (and thus these
would correspond respectively to B and A in Fig.~1(c)), and the larger lobe field would be formed by B2 
and the negative field north and east of B2 (these would correspond respectively to B and C in Fig.~1(c)).

Which, if either, of these explanation explains the strong intensity of the loops west of the B1---B2 neutral
line would require more investigation to determine with confidence; those investigations are beyond 
the scope of the present study.

We present one more example where we can identify an erupting minifilament, this time for an eruption
occurring in region R3 of Figure~3. \iris\ data are not available for this event.   We do not present
separate AIA high-cadence movies of this event, but from the video accompanying Figure~4(c), over 
January~14 10:22~UT---10:36~UT there is a minifilament (of width $\sim$5$''$) moving in the southwest
direction away from the neutral line north and east of positive magnetic clump labeled B7 in Figure~6(d). 
Subsequently this minifilament erupts outward, apparent in the video from 10:36~UT  to 10:42~UT\@.  From
Figure~4(d), and its accompanying video at 10:38~UT, we see brightenings at the approximate location of
the neutral line from where the minifilament erupted.  Again the brightest location is shifted somewhat,
in this case to the southern portion of the neutral line  where the fields along that neutral line are
strongest; but the brightening also extends further west of that neutral line, on top of the central
portion of the B7 positive-polarity clump and negative field to its south.   Similar brightenings are
visible in X-rays in the XRT video accompanying Figure~3, over 10:36~UT---10:37~UT\@.  Subsequently, the
western brightening becomes extremely  intense (apparent after saturation has subsided in the Fig.~3 XRT
video at 10:53:46~UT\@).  This again could be a very bright loop from the external reconnection, rooted in
strong field.

\section{Magnetic Field Evolution}
\label{sec-b evolution}

From the above observations, apparently the factor responsible for the jetting occurring in the regions
labeled R1, R2, and R3 in Figures~3(a) and 4(c) is the presence of the minority-polarity positive
regions in the overall dominant-polarity negative field in the northwest portion of the active region. 
This is in accord with the general picture for jets presented in the schematic of Figure~1.  Next we
examine the evolution of the magnetic flux patches over a period of jetting in subregion R2.

From the video accompanying Figure~6(b), we see that rapid evolutionary changes, including flux emergence
and flux cancelation, are occurring throughout the observation period.  In order to make tractable
investigation of the relationship  between the magnetic changes and the jets, we limit our focus to the
period  when jetting is concentrated in region R2 of Figure~3; based on, e.g., the Figure~6 video, we see
that the time window on 2015 January~14 over 12:35~UT---22:35~UT contains most of these R2 jets. 

Figures~8(a) and~8(b) display the magnetic evolution over the time period, where the top two panels
corresponding to the two parallel rectangular boxes in Figure~6(b), with Figures~8(a) and~8(b)
respectively corresponding to boxes~(b1) and~(b2) in Figure~6(b).  Figures~8(a) and~8(b) show 
distance-time plots, where the vertical-axis distance runs along the $x'$ axis in Figure~6(b), and the
plotted quantity is the sum of the signed flux in the $y'$ direction: $\int B\diff y'\label{eq:1}.$  These
Figure~6(b) boxes were selected and oriented so that we could follow the evolution at the source neutral
line for the jets  of region R2 of Figure~3, and also so that the $x'$ direction would approximately be
aligned with the drift direction of the positive-polarity clumps.  We selected these two boxes rather than
a single box covering the full area of b1+b2, because the positive flux of this R2 area approximately
splits into two different clumps, each approximately contained in one of these two boxes, over the time
period of Figure~8; one can see this from the animation accompanying Figure~6(b),  which shows these two
boxes overlaid over the time period of Figure~8. It is nonetheless hard to isolate the white flux
completely over the full time range of Figure~8 and still show the features that we wish to highlight. 
Thus there is some drift of positive flux into box~(b1) over 16:30~UT---18:30~UT; this increases the value
of the positive flux contribution to the integral over $\diff y'$, and this accounts for a modest increase
in the white intensity in the Figure~8(b) plot over this time.

Figure~8(c) shows the integrated intensity as a function of time of the blue box of Figure~6(b), and shows
peaks at times of jet brightenings originating from region R2 of Figure~2.  There is a cluster of jetting
between about 13:45~UT and 15:45~UT, and a weaker cluster between about 20:00~UT and 22:30~UT, with the
times indicated by the blue and magenta line pairs; these clusters of jets at the R2 location are
apparent in the movies accompanying Figure~4. Both of these cluster times correspond to times of
unmistakable flux cancelation at the main neutral line of region R2 in Figure~4(c), as can be seen by
referencing the colored-line pairs in Figures~8(b).

A strong jet near 19:00~UT in Figure~8(c) corresponds to cancelation visible in Figure~8(a), highlighted
by pairs of orange lines in both panels.  This cancelation does not stand out as much as those in
Figure~8(b); examining the movie accompanying Figure~6(b) shows that this is because, even though flux
cancelation occurs at the neutral line, a strong positive flux element still remains immediately adjacent
to the canceling neutral line.  In summary, all of the main jets of this R2 subregion occur at the sites
of flux cancelation.

Figure~9 shows the change with time of the total positive-polarity flux in each of the two boxes of
Figure~6(b).  Since some positive flux does enter into the box with time, we cannot trust in detail the flux
changes over all times, due to the aforementioned positive-flux drifts into the boxes; but this effect seems
to be small even over the time of the most obvious such drift into the box, 16:30~UT---18:30~UT\@.  From the
movie accompanying Figure~6(b), drifts of positive flux out of the boxes~(b1) and~(b2) appear to be even less
extreme.   Cancelation in box~(b1) corresponds to the times of enhanced jetting activity over approximately
14:00---16:00~UT (blue lines in Figs.~8 and~9) and over approximately 21:00---23:30~UT (magenta lines), and
cancelation  in box~(b2) corresponds to the jet near 19:00~UT (orange  lines).  (Although prominent in AIA
images, that 19:00~UT jet was not observed by XRT due to a gap in  the observations.) Therefore these plots
confirm our interpretation above that flux cancelation is taking  place near the times of the jets.

We see in Figure~9(a) a rate of flux decrease of  $\sim$0.5$\times
10^{19}$~Mx~hr$^{-1}$ for about an hour around the time period where there is one outstanding jet in the AIA
94~\AA\ intensity plot of Figure~8(c).  For the period over 14:00---15:00~UT in Figure~9(b) the drop rate is
$\sim$1.5$\times 10^{19}$~Mx~hr$^{-1}$, and from Figure~8(c) we see that over that same time period
(14:00---15:00~UT) there are about three main intensity peaks.  (Especially for this 14:00---15:00~UT time 
window, we suspect that some cancelation is still occurring at the times when jets occur just before or 
after the times where the flux is dropping in Fig.~9(b),  but that flux drop is 
not well-captured by the restricted box region of Fig.~6(b) we used to make that flux plot. This is why 
the blue lines in Figs.~8 and~9 cover a wider time window than just 14:00---15:00~UT for these jets.)   From the 
videos accompanying Figures~4 and~6, the smaller peaks of Figure~8(c) over this time period are due to weaker jet
events and/or secondary events.  For the $\sim$21:00---22:30~UT period, there is a drop in flux with a rate
of  $\sim$1.8$\times 10^{19}$~Mx~hr$^{-1}$, and over this time there are about three main jetting 
episodes identifiable in the Figure~8(c) light curve.   Therefore we can say that the flux cancelation rates
range over (0.5---1.8)$\times 10^{19}$~Mx~hr$^{-1}$ for the active region jets that we observe over
this R2 region over the period we examined.  The average cancelation rate weighted by the number of jets at
each of the above rates is $(1.5\pm 0.5) \times 10^{19}$~Mx~hr$^{-1}$, based on an average of about 
seven violent active region jets.

From this, we can estimate the amount of flux going into the buildup to each of these stronger jets (i.e., the jet 
activity corresponding to the three peaks in the Fig.~8(c) light curve).  For the three 
such jets of the 14:00---15:00~UT 1-hr time window, this averages to $\sim$5$\times 10^{18}$~Mx.  Similarly,
for the three such jets the 21:00---22:30~UT 1.5-hr time period, the average is $\sim$4$\times 10^{18}$~Mx, and
for the single jet of the $\sim$1-hr time period of the Figure~9(a) jet the average is also 
$\sim$(4---5)$\times 10^{18}$~Mx.  So we can say that roughly $\sim$5$\times 10^{18}$~Mx of flux cancels 
prior to each of these stronger jet episodes.  At least for  subregion R2, upon which we performed the detailed 
analysis
above, the jetting episodes continue until all of  the clump of minority flux disappears.

\section{Energetics}
\label{sec-energetics}

From the above-estimated amount of flux canceled prior to each of the jet episodes, 
we can make a rough estimate of the amount of energy in our active region violent jets,
assuming that they are produced via erupting flux ropes as in the Figure~1 schematic. 

We begin by estimating the size of the flux rope.  We have already noted that the width of the erupting
filament strands is $\ltsim 2''$; so let's take a width $2r$ of $2''$.  Figure~6(d) shows the length of the B3---B4
neutral line to be $\sim$10$''$; we take this to be the length, $l,$ of the flux rope, resulting in a ratio
$l/r\sim 10$ for the flux rope.  From this, we can derive a cross-sectional area, $A,$ of the flux rope to be
$\pi r^2 \sim 1.7 \times $10$^{16}$~cm$^2$, and volume, $V,$ to be $\pi r^2l\sim$1.2$\times 10^{25}$~cm$^3$.

We estimate the magnetic field strength, $B,$ of the flux rope by assuming that the
total flux flux accumulated for each jet calculated above, 5$\times 10^{18}$~Mx, flows through the area of 
the flux rope, so that the total flux is $BA$\@.  Taking that area to be the value $A$ we estimated for 
the flux rope cross section, this yields $B\sim 300$~G\@. This might be considered a rough 
estimate for $B$, considering that the flux rope would be expected to form nearer to the photosphere than at the time we
observe erupting strands in the AIA images, and hence it would be more compact than what we calculated above, 
reducing $A$ and increasing $B$ somewhat; and on the other hand the true flux rope may have a cross-sectional area
larger than that of the strand, for example if the strand has an enveloping cavity around it, reducing $B$ somewhat.  
For an approximate upper limit for $B$, we use the value of the positive-polarity
magnetic flux clumps that we measure from the magnetograms, such as B4 in Figure~6(d), where we find maximum
values in the range 300~G---500~G\@.  Thus we can consider a range for $B$ of the flux rope to be 
approximately $\ltsim 300$---500~G\@.  

Using $B=300$~G gives a magnetic energy in the flux rope 
$E_{fr}=V(B^2/8\pi) \approx 4.3 \times 10^{28}$~erg.  Using our upper estimate for $B$ would increase this
by a factor of 3.  Keeping in mind our caveats for our $B$ values, we estimate the energy contained in the 
erupting-minifilament flux rope of the
Figure~1 schematic to be $\sim$10$^{28}$~erg~---~10$^{29}$~erg for the violent jets.  Deductions from observations
of jet energy values range over 
$10^{25}$~erg~---$10^{29}$~erg \citep[e.g.][]{shibata_et92,shimojo_et00,pucci_et13,raouafi_et16}, and so our
derived estimate of flux rope energies is sufficient even for the greatest of these.

\section{Summary and Discussion}
\label{sec-discussion}

We have examined a series of active region coronal jets, bright in X-rays and fast-evolving
(``violent''), like the ones that were difficult to analyze in Paper~1.  We find that the jets tend to occur in
localized regions, which we initially identified in XRT soft X-ray images and which we have labeled R1,
R2, and R3 in Figure~3.  Magnetograms show that each of these regions is broadly consistent with the
schematic of Figure~1, where in each of R1, R2, and R3 there is a positive-polarity island butting up
against a negative-polarity region, within a generally-negative-polarity larger region; that is, the 
positive and negative polarities are respectively the minority and majority polarities in the overall 
region.  Figure~1 was initially drawn to describe jets in coronal-hole regions \citep{sterling_et15}, 
but recently we have found that it applies to jets in quiet Sun regions \citep{panesar_et16a}, and also to at
least relatively-slowly-developing active-region coronal jets (Paper~1).  Here we confirm that the same
general picture also applies to at least many of the violent active region jets.  We also found
that cancelation occurs at the neutral line on which the jet activity commences, similar to what has
been found in several other on-disk observations of coronal jets.  In comparison with some other jets 
however, when detectable, minifilaments 
often appear as narrower ($\ltsim 2''$)
and harder-to-see strands than our earlier-observed erupting minifilaments in jets in coronal holes
and quiet regions.  

We also found that loops
neighboring the JBP can be more intense than the JBP.  This phenomenon has however also been seen in some 
coronal hole jets \citep{moore_et13}, and therefore is not limited to the violent jets studied here.  How
this fits with the picture of the Figure~1 schematic (e.g.\ whether they form the JBP for a secondary 
jet) is a topic for later consideration.

In Figure~6 we examined in detail the set of jets occurring in the R2 region.  High-cadence AIA EUV
images and IRIS UV images overlaid with HMI magnetograms corroborate that the jets start with
brightenings roughly along a strong-gradient neutral line between the minority and majority fluxes,
that the minority-polarity region is linked by bright-loop extensions to farther-away majority-polarity
regions,  that the jet forms above this mixed-polarity setup, and that the jet spire drifts with time
away from the location of the initial neutral-line brightenings; all of these properties are consistent
with the minifilament-eruption schematic for jet production (Fig.~1).  

For this same series of jets in region R2, we follow the magnetic evolution in Figures~8 and~9.  We find
that the jet occurrences cluster around times when the most-prominent flux cancelation is ongoing, and 
there are few strong jets (meaning those showing in the AIA 94~\AA\ lightcurve of Fig.~8(c)) outside  of
those cancelation times.  Thus, our results are consistent with the jets resulting from magnetic flux
cancelation: when the strong cancelation episodes stop, the strong jetting stops.  This is also consistent with
recent findings for individual jets \citep[e.g.][]{hong_et11,
huang_et12,adams_et14,young_et14a,young_et14b,zhang_et14}, and with recent findings examining multiple
jets \citep[e.g.,][]{chen_et15a,chen_et15b,panesar_et16a}.   Other studies \citep[e.g.][]{innes_et13}
observe jet-like phenomena originating from sites of network-field cancelation.   \citet{shelton_et15} 
report jets resulting from flux emergence, although the flux changes provided in their figures also show flux 
decreases near the time of jet brighting, and so flux cancelation  as the source of the jets in that
case is not ruled out.


In some cases, jets do occur from locations of flux emergence.  But for at least many of these cases,
close examination shows that it is the cancelation of one pole of this emerging bipole with a nearby
majority-polarity flux clump that results in the jet \citep[e.g.][]{shen_et12}. Our active regions events
here, and also those of Paper~1, also contain cases where flux emergence occurs, but as with
\citet[][]{shen_et12} the jets originate from the cancelation sites.  In active regions especially, it is
often difficult to separate emergence from cancelation, since both occur frequently during some times of
the region's life.  In quiet Sun regions, cancelation consistently seems to be the source of the  jets
\citep{panesar_et16a}, even though in most cases the fluxes involved are long past their emergence phase. 
This leads us to suspect that, at least in many cases, {\it cancelation is the fundamental
magnetoconvection process responsible for the pre-eruption buildup and triggering of jets,} even in active
regions.   It is conceivable that cancelations at much smaller
rates at the  edges of supergranular networks might produce even smaller scale jets, such as spicules
\citep{sterling_moore16}.




As mentioned above, compared to many of the violent jets in active regions, minifilaments were easier to
identify in the quiet Sun, coronal hole, and less-violent active-region-jet cases.  For example, from the figures
and animations  of \citet{panesar_et16a}, the quiet-Sun minifilament widths near their time of eruption onset 
are $\sim$2$''$---3$''$, and from the figures and animations of Paper~1 the  minifilaments leading to the weaker
coronal jets (but producing surges/sprays) have widths of $\sim$3$''$---4$''$.  Here we have found
harder-to-detect strands of what appears to be cool filament-like material of widths $\ltsim 2''$ in the
eruption that drives the jet, akin to the minifilament eruptions in the less-violent jet eruptions.    And in fact,
with our new insights into observing these minifilament strands, we are also able to identify similar strands in
the data for some of the violent jets of Paper~1.

\citet{schmieder_et13} also report observations of many threads along an active region jet,
which may be the same features that we observe.  They \citep{schmieder_et13} suggest that the strands may not  be a
fully-erupting ``flux rope.''  Under this view, it may be that strands of a  lower-laying minifilament structure
``peel off'' as strands, and erupt to form part of the jet.  In our view this is plausible, whereby the peeling off
would be triggered, somehow, by the flux cancelation that, from Figures~8 and~9, is clearly cotemporal with bursts of 
jet activity.  


We found that observing these minifilaments/strands in active region jets is much more difficult than
observing minifilaments in quiet Sun jets that we examined in \citet{panesar_et16a}.  As suggested by
Figure~7, even in relatively clear cases where we can see the  erupting minifilament strands prior to
jet-spire production, they can be hard to see without zooming in on the jet-base region and using
high-cadence images.  

Even in cases where we do not  observe a minifilament or strand in the active region jets, we suspect that
the same (Fig.~1) process occurs, but the absorbing material is not detected because of one or more of these
reasons: (a) Its width is too small for us to resolve. (b) It is obscured by surrounding dark material. (c) It
is surrounded by bright material resulting from internal reconnection of the legs of the field enveloping the
erupting minifilament/strand, producing a bright ``cocoon''  \citep[][]{sterling_et11} wrapping around the
erupting minifilament.  Similarly, external reconnection between the minifilament/strand-enveloping  field and
the surrounding (relatively strong) ambient active region coronal field might produce a spire that is
particularly bright (shaded regions of Figs.~1(c) and~(d)), obscuring or partially obscuring the cool material. (d) The 
eruption may be occurring without cool
material being ejected on the erupting field, just as some large-scale flux-rope eruptions occur without an
obvious cool filament (and some of this bright material may result from brightening of the erupting
(mini)filament material itself; see, e.g.\ \citeauthor{su_et15}~\citeyear{su_et15}). (e) A corollary to  (d) is
that, because the evolution of the magnetic fields is relatively fast in active regions compared to non-active
regions, there may not be time for enough cool  material to accumulate on a newly-formed flux rope for a
substantial minifilament to develop.  We do not know how quickly a (mini)filament can develop on a flux rope,
but at least in the case of non-active-region filaments it can take a number of hours
\citep[e.g.,][]{berger_et12}.

Regarding point (c), our observations do show that in the 304~\AA\ images there is a mixture of brighter and
darker strand-like features, and this is consistent with the minifilament material being partially obscured by
brighter material during jet onset in some cases, and with that darker material enveloped
by a brighter shell of material that has undergone reconnection and/or a bright spire.


Observations by \citet{hong_et16} using the New Vacuum Solar Telescope (NVST) show high-resolution
\halpha\ images  of an active region jet location reveal a very narrow (width $\sim$1$''$) minifilament
that erupts to form a jet similar to those observed here
(cf.~\citeauthor{panesar_et16b}~\citeyear{panesar_et16b}),  in agreement with the process described in
\citet{sterling_et15} (Fig.~1).  So erupting minifilaments of this size could be the erupting strands we observe. 
Also, observations by \citet{li_et15} of an active region jet show that it results when a filament,
clearly visible in the NVST \halpha\ images, rises up and apparently reconnects with open field, with some
of the filament material flowing out along the newly-open field lines.  This process also could be
responsible for the source of the erupting minifilament strands that we observe here.  Sorting through
these possibilities however is not possible for the cases we present here with the data sets we use.

In conclusion: We observed X-ray coronal jets at the leading edge of an AR\@.  The jets were
in primarily three localized locations of the region. Those regions were where flux cancelation occurred,
and the jets recurred at those respective locations until all of a clump of minority flux canceled and
disappeared.  The overall magnetic setup of all the observed jets agreed with that of the schematic presented
by \cite{sterling_et15}, which was originally drawn to describe coronal jets observed in polar coronal holes, where
erupting minifilaments were found to be the source of the jets.  There are however some differences from the
situation with those coronal hole jets, as well as with coronal jets observed in quiet regions \citep[which][found 
to behave in a fashion similar to the coronal hole jets]{panesar_et16a}: In the case of AR jets 
\citep[both here and in the previous study of AR jets in][]{sterling_et16}, clear minifilaments were 
frequently not apparent, and also the JBP's intensity was frequently surpassed by brightenings from different
enhanced-polarity locations of the jet's base.  We have put forth some ideas for why these differences exist.
Our educated guess is that most AR jets work in the same manner as many non-AR jets, with apparent 
differences resulting from the environmental circumstances (such as stronger
magnetic fields) or secondary eruptions/jets.  Ultimately, further studies are needed 
to determine with certainty whether all coronal jets in all solar regions result from basically
the same mechanism,
or if instead there is a fundamental difference in the way jets of the different solar regions can
be produced.

\acknowledgments

We thank K. Reeves for directing us to the data set we examined, and we thank J. Klimchuk and E. DeLuca for useful
discussions.  A.C.S., R.L.M., and D.A.F. were supported by funding from the Heliophysics Division of NASA's Science Mission
Directorate through the Heliophysics Guest Investigators (HGI) Program, and the \hinode\ project.  N.K.P's research was
supported by an appointment to the NASA Postdoctoral Program at NASA MSFC, administered by Universities Space Research
Association under contract with NASA\@. Hinode is a Japanese mission developed and launched by ISAS/JAXA, with NAOJ as
domestic partner and NASA and STFC (UK) as international partners, and operated by these agencies in co-operation with ESA
and NSC (Norway). FM was supported by NSF's Research Experience for Undergraduates (REU) Program.

\clearpage

\figcaption{Schematic showing jet generation via minifilament eruption, as proposed in
\citet{sterling_et15}.  This schematic is a modified version of that in \citet{sterling_et16} (Paper~1). 
(a) A minority-polarity (positive in this example) flux clump sits inside of a majority-polarity
(negative) ambient background field.  The positive polarity has connections to immediately-surrounding
negative polarity field, naturally forming an anemone-shaped field  pattern \citep{shibata_et94} in 3D;
these frames show a 2D cross section of the anemone.  Here, the left side of the anemone contains a
sheared, non-potential field; thus in 2D the structure consists of a small lobe adjacent to a larger lobe
in this rendition (situations on the Sun might change the geometric ratios while maintaining the overall
topology). In the picture, cool filament material (blue circle) gathers in the sheared field above and
along this neutral line, analogous to filament formation on larger-scale sheared-field neutral lines; this
is the minifilament.  (b) Initiated by some (unspecified) trigger mechanism, the minifilament-carrying
field (flux rope plus envelope) erupts outward, and travels between the larger lobe and the ambient
field.  This leads to magnetic reconnection (red X) interior to the erupting minifilament field ({\it
internal} reconnection), making a flare loop (arcade in 3D) lower in the atmosphere (red simicircle); we
identify this flare arcade with the observed JBP\@.  (c) Upon reaching the field on the far side of the
large lobe, the envelope of the erupting minifilament field reconnects with the ambient field
(right-hand-side red X), making a new open field line (red open field line), and a new loop over the large
lobe (large semicircular loop); the plasma on both of these newly-reconfigured field components is
energized via heating and particle acceleration by the reconnection, and hence represented by the red
color.  (This is {\it external reconnection,} since it occurs external to the minifilament-flux-rope field
that is driving the reconnection.  This is also called interchange reconnection, e.g.\
\citeauthor{crooker_et02}~\citeyear{crooker_et02}.) Plasma pressure and magnetic force result in
acceleration of hot plasma along the newly-opened field (orange shadings); we identify this heated plasma
with the observed jet spire.  In this frame, so far only the outer envelope of the erupting minifilament
flux rope has undergone this reconnection, but the field threading the cool minifilament material has not,
and so only a hot jet occurs so far.  (d) As  the minifilament eruption continues, the external
reconnection can eat through the envelope field surrounding the cool minifilament material and the open 
field that threads that minifilament material.  This cool
material can then escape along that new open field, resulting is a cool (e.g., visible in AIA 304~\AA\
images) jet, along side of or intertwined within the hot jet.  Labels  ``A,'' ``B,'' and ``C'' are
referred to in the text to compare with magnetic polarities of observed jetting regions.  \label{fig1}}

\figcaption{Overview of the jet-productive active region NOAA AR~12259, in \sdo/HMI (a) white-light intensity, and 
(b) magnetogram images.  In the magnetogram, black and white represent negative and positive polarities,
respectively, with values greater (less) than 1000~G (-1000~G) saturated. The boxed region of (a) is a 
jet-prolific location over the period of \hinode/XRT observations over
2016 January 13---14; jets of this study are concentrated in this region, and this boxed region is the field of
view  (FOV) of Fig.~4 below.  For these, and for all other solar images in this paper, north is up and  west is to the
right, and the images have been differentially rotated to the common time of 2016  Jan~14 16:00~UT\@ . 
\label{fig2}}

\figcaption{\hinode/XRT X-ray images of the boxed region of Fig.~2(a), with R1, R2, and R3 showing 
X-ray-bright subregions where multiple jets were concentrated over our observing period.   Two different,
but essentially homologous jets are occurring in subregion R2 in (a) and (b).  Accompanying animations
show that numerous jets occur over the XRT observing period.  XRT was not observing this region over the 
time gap between 2016 Jan~13 17:44 and Jan~14 6:16 UT in the animation.  (The abscissa in  the animation
prior to the time jump was artificially shifted to reflect the 2016 January 14 16:00~UT coordinate location;
this was necessary because that location was outside of the XRT FOV at those observations times.)
\label{fig3}}

\figcaption{AIA images of the jetting regions in (a) 193~\AA\ and (b) 94~\AA, and respectively the same
in (c) and (d) but with HMI magnetogram contours overlaid (with overlay times given in parentheses in the title).
Contour
levels are at $\pm$30, 50, 100, 750, and 1400~G, where red and green represent positive and negative
polarities, respectively. Labels R1, R2, and R3 are above (north of) the neutral lines of the
correspondingly-labeled regions in Fig.~3.  Animations of these four panels  are available.
\label{fig4}}

\figcaption{Light curve of the AIA 94~\AA\ intensity, integrated over the FOV of Figs.~3 and~4, at 2~min 
cadence.  Peaks show incidences
of flaring, often accompanied by jets, usually emanating from one or more of the subregions labeled R1, R2, and R3
in Fig.~3.   \label{fig5}}

\figcaption{(a) AIA 171~\AA\ image of a jet examined in \S\ref{sec-r2}, which was one of several
homologous jets occurring in region R2 of Fig.~3.  The tip of the black arrow points to a faintly visible
erupting minifilament strand. (b) An HMI magnetogram of the region of (a), with values greater (less) than 
300~G (-300~G) saturated.  The dashed lines  show the
north and west edges of the FOV of the \iris\ images in the subsequent panels.   Boxes (b1) and (b2)
inside of panel~(b) show regions used to track magnetic development in Figs.~8 and~9, with  the
($x'$,$y'$) coordinate system for those two boxes as indicated.  The blue rectangle shows the region over
which intensities in Fig.~8(c) are determined.  (c) \iris\ 1400~\AA\ slit-jaw image of the jet of (a); the
FOV is different from  that of (a) and (b).  (d---f)  Same as (c), but with HMI magnetograms overlaid. 
The title of each of these three panels gives the time of the 1400~\AA\ greyscale image, and in parentheses, 
the time of the magnetogram.  Labels in (d) identify magnetic flux clumps, and are referred to in the text; a
flux clump labeled B1 appears in Fig.~7, but that clump had disappeared by the time of this image. 
Animations of  panels (a), (b), (c), and (d---f) are available.  The animation corresponding to panel~(b)
runs twice, the first time with only the magnetogram, and the second time with the boxes~(b1) and~(b2)
overlaid for the time period corresponding to  the plots in Fig.~9. \label{fig6}}

\figcaption{AIA 193~\AA\ images showing minifilament ``strands'' at the start of a pair of jets
from  region R1 of Fig.~3, where (a) and (b) show the weaker of the two jets, and (c) and (d) show the
stronger  of the two jets.  Arrows in (a) and (c) show the minifilament prior to eruption, at times when
they are  apparently ``crawling up'' a magnetic dome that has enhanced emission, corresponding to the
large bipole in the Fig.~1 schematic.  In (d), magnetogram contours, with properties of that in Fig.~4,
are overlaid onto the region (with times given in parentheses in the title).  Labels in (d) are as those
in Fig.~4(c); region B1 appears here, but has  disappeared by the time of Fig.~4. High cadence (12~s)
animations of panels (a---c) and panel~(d) are available.
\label{fig7}}

\figcaption{Magnetic evolution over the times of the R2 region (labeled in Fig.~3) jetting.  (a)
Evolution in time of the  magnetic elements in rectangular box labeled ``(b2)'' in Fig.~6(a), where the
ordinate is distance along the $x'$  axis of the rectangular box, and the abscissa is the net signed magnetic
flux from integration along the $y'$ direction of that  rectangular box.  (b) Same as (a), but for the
rectangular box labeled ``(b1)'' in Fig.~6(b).  (c) Light curve of the AIA 94~\AA\ intensity, integrated
over the blue box in Fig.~6(b), with 2~min cadence; peaks show the times of substantial jetting activity 
from region R2.  Pairs of same-colored lines in (c) highlight times of two clusters of jets (blue and 
magenta), and of a strong single jet (orange).  These correspond respectively to periods of flux cancelation 
in (b) and (a).
\label{fig8}}

\figcaption{Magnetic flux changes with time, of the positive-flux clumps contained in the regions 
marked (a) box~(b2), and (b) box~(b1) in Fig.~6(b).  Sharp drops in flux coincide with periods of
increased jetting, as can be seen by comparing the identically-colored line pairs of Fig.~8(c) with 
the line pairs in this figure.
\label{fig9}}

\clearpage
\pagestyle{empty}
\begin{figure}
\plotone{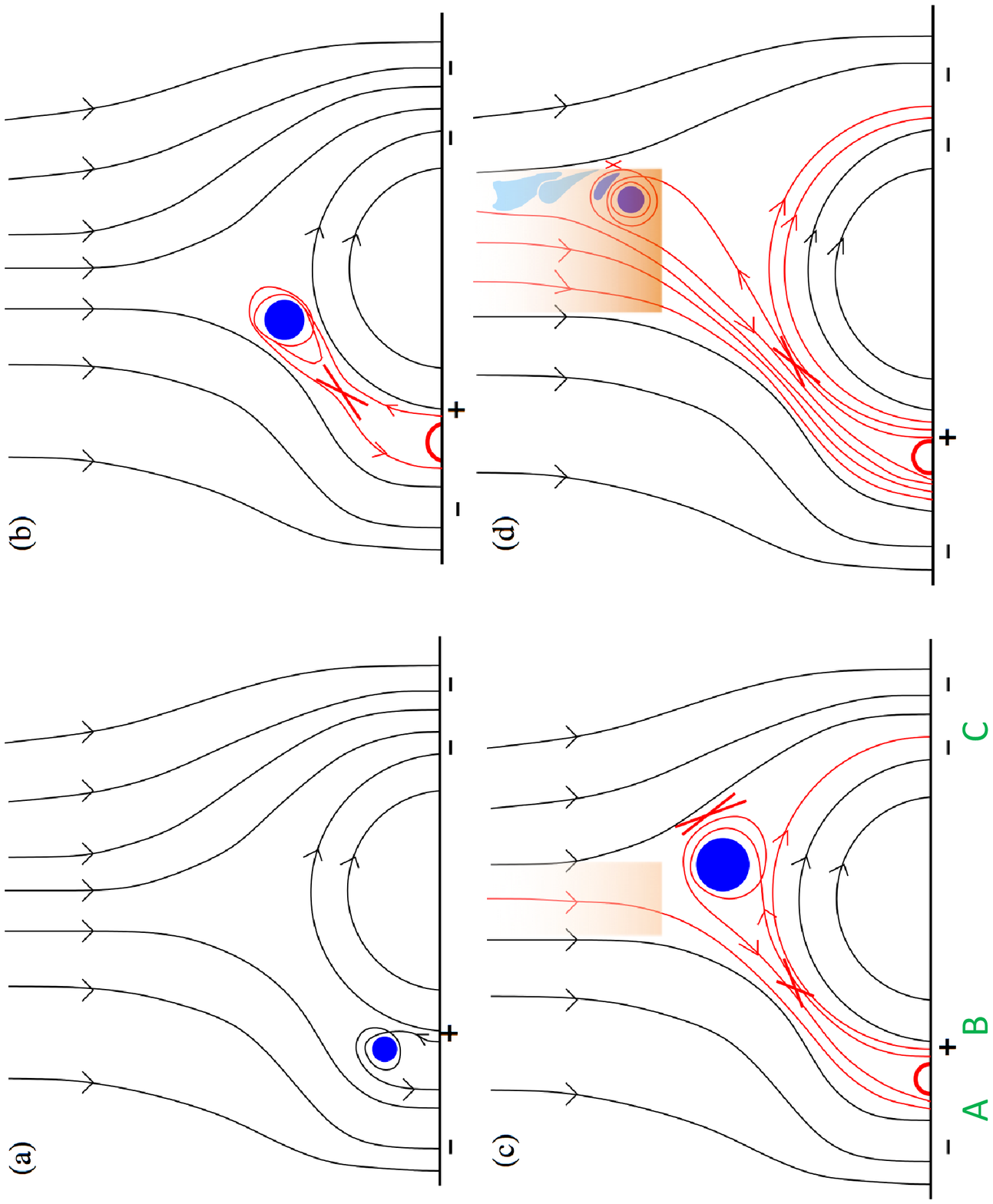}
\centerline{Figure~1}
\end{figure}
\clearpage

\begin{figure}
\hspace*{-2cm}\includegraphics[angle=0,scale=1.1]{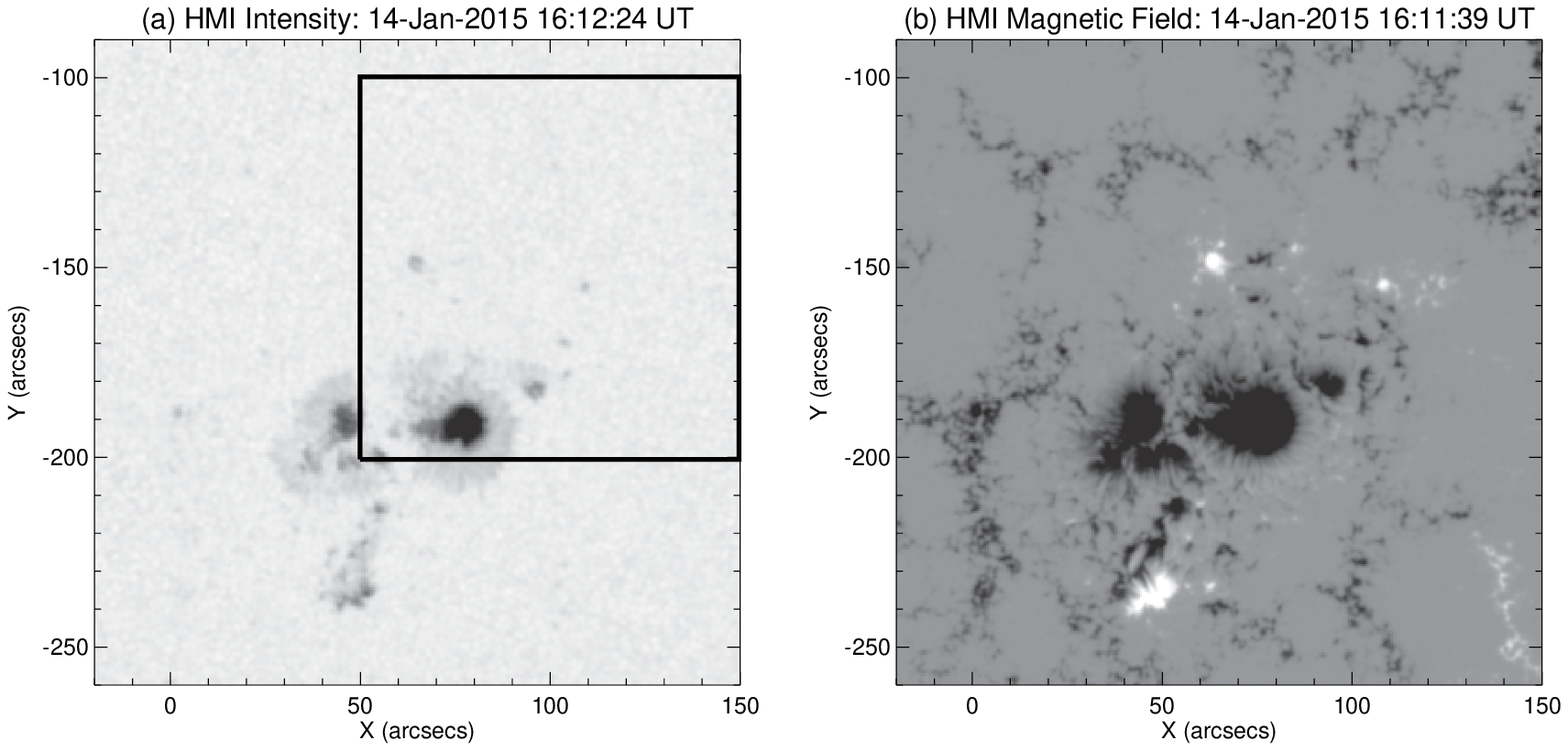}
\centerline{Figure~2}
\end{figure}
\clearpage

\clearpage
\pagestyle{empty}
\begin{figure}
\plotone{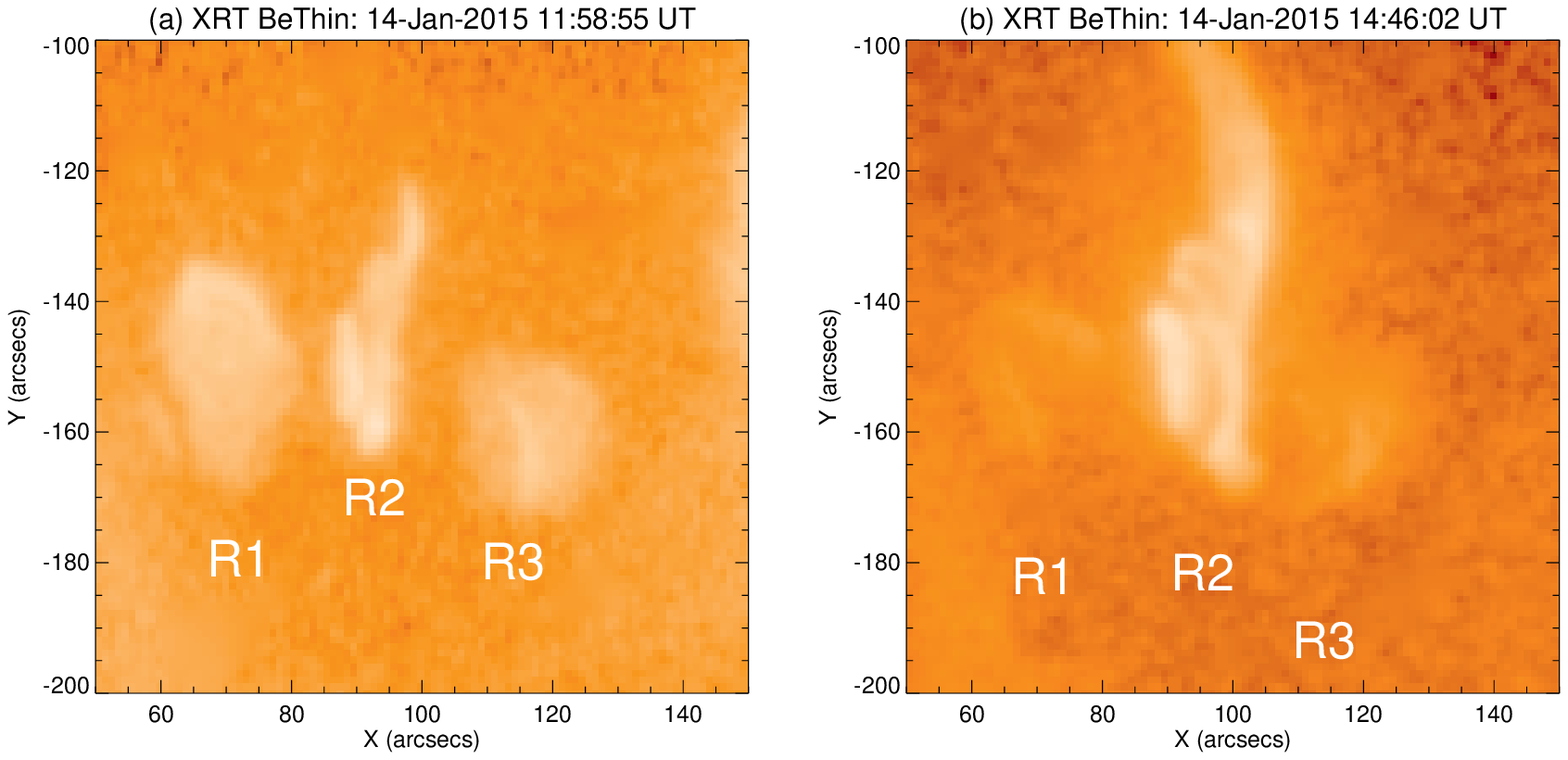}
\centerline{Figure~3}
\end{figure}
\clearpage

\clearpage
\pagestyle{empty}
\begin{figure}
\plotone{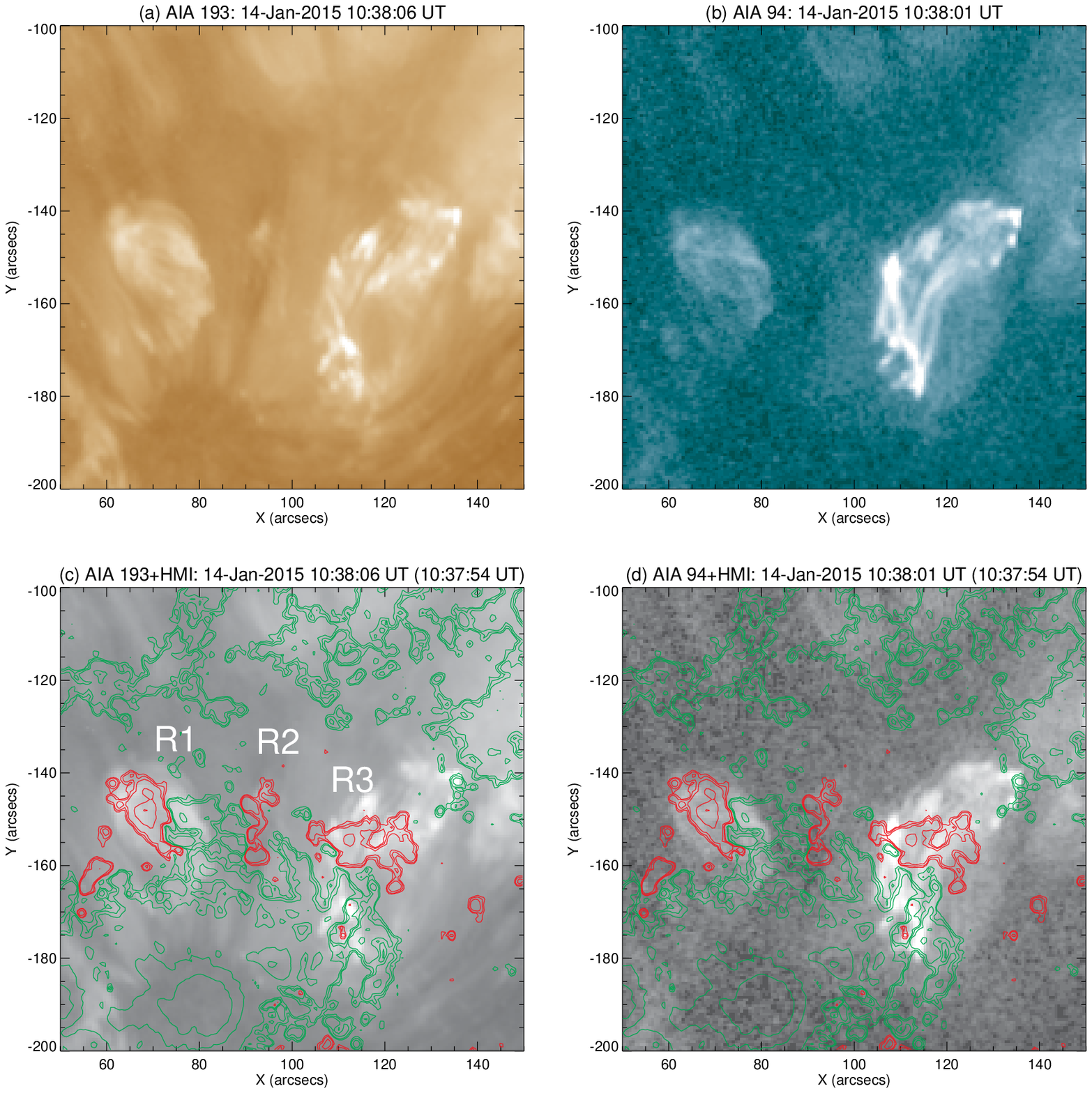}
\centerline{Figure~4}
\end{figure}
\clearpage

\clearpage
\pagestyle{empty}
\begin{figure}
\plotone{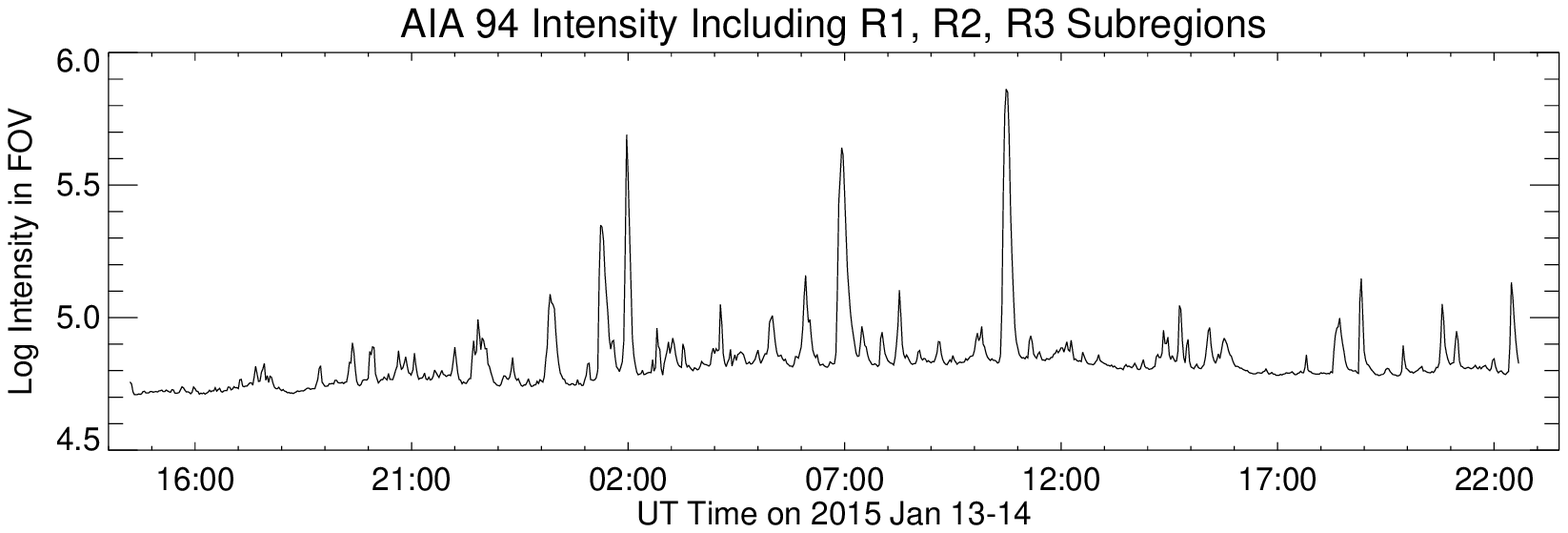}
\centerline{Figure~5}
\end{figure}
\clearpage

\clearpage
\pagestyle{empty}
\begin{figure}
\plotone{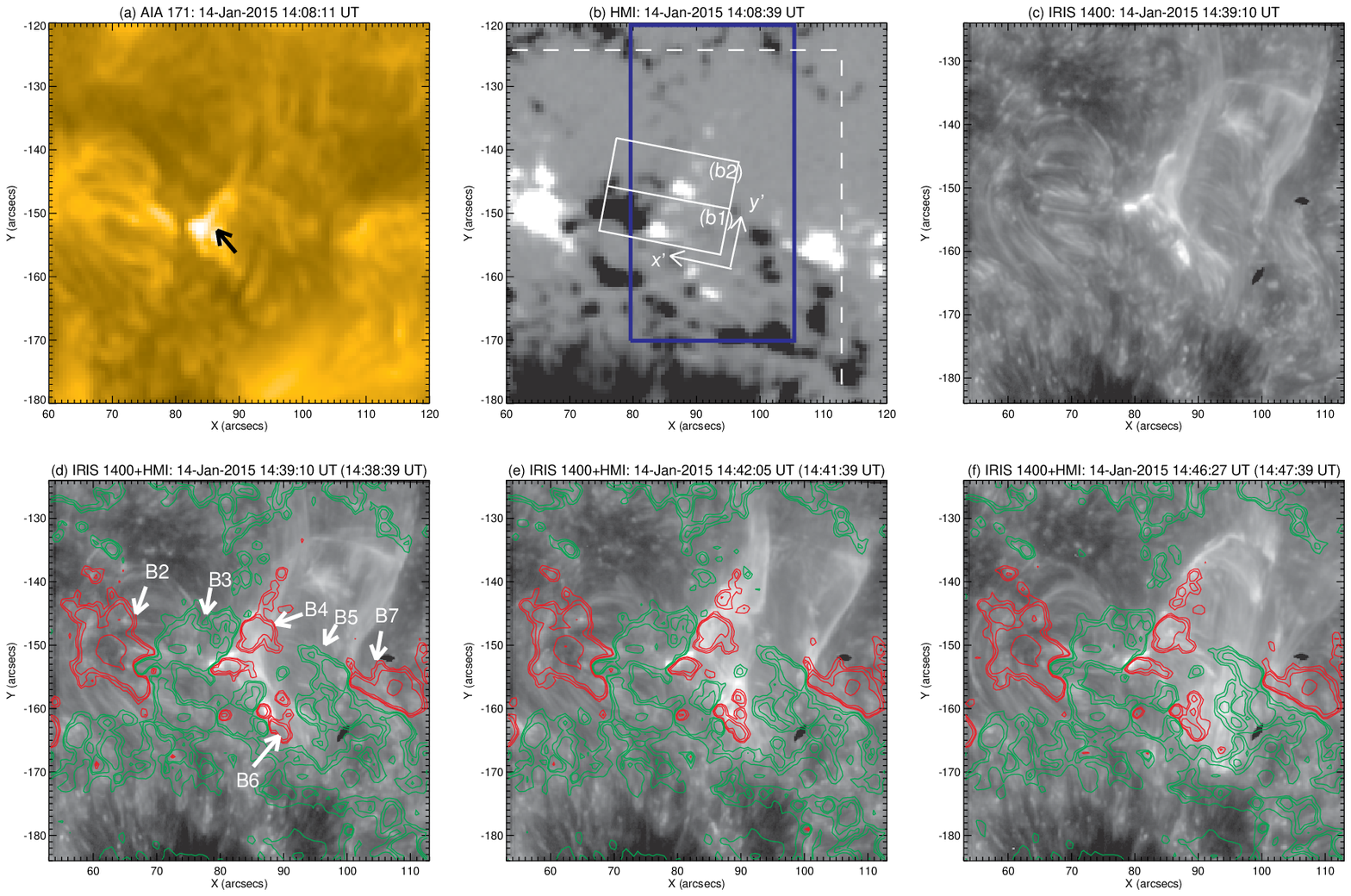}
\centerline{Figure~6}
\end{figure}
\clearpage

\clearpage
\pagestyle{empty}
\begin{figure}
\plotone{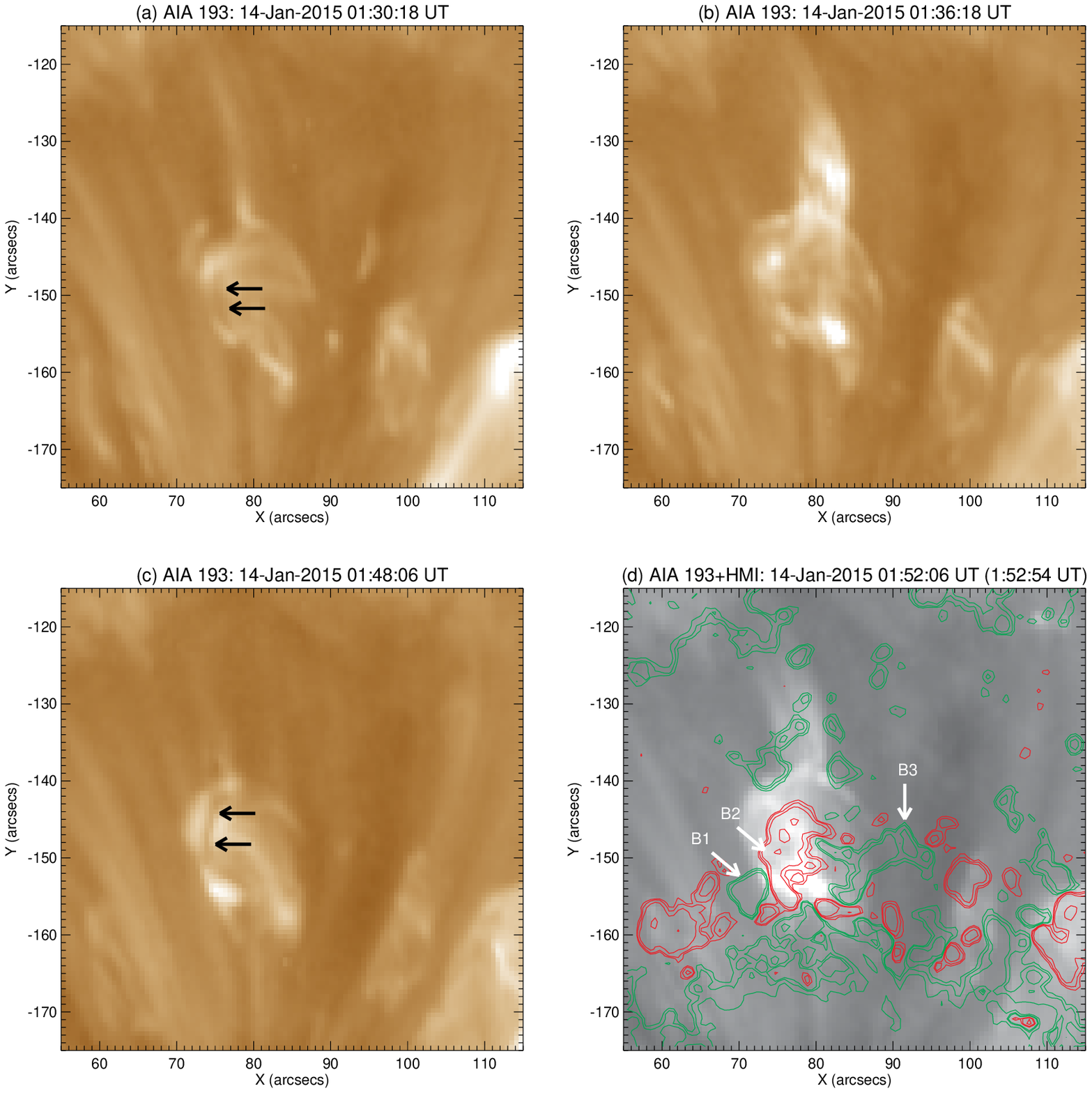}
\centerline{Figure~7}
\end{figure}
\clearpage

\clearpage
\pagestyle{empty}
\begin{figure}
\plotone{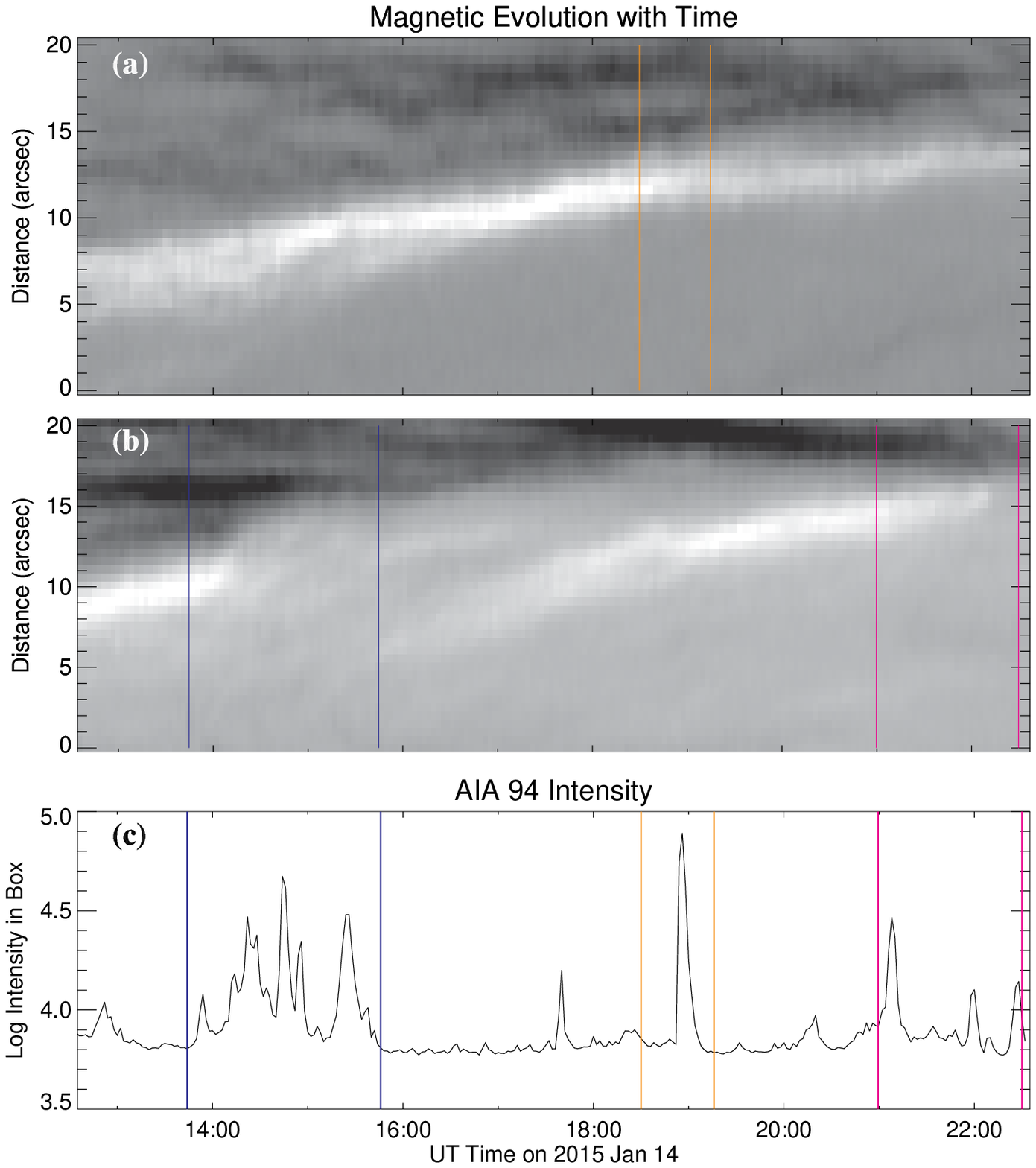}
\centerline{Figure~8}
\end{figure}
\clearpage

\clearpage
\pagestyle{empty}
\begin{figure}
\plotone{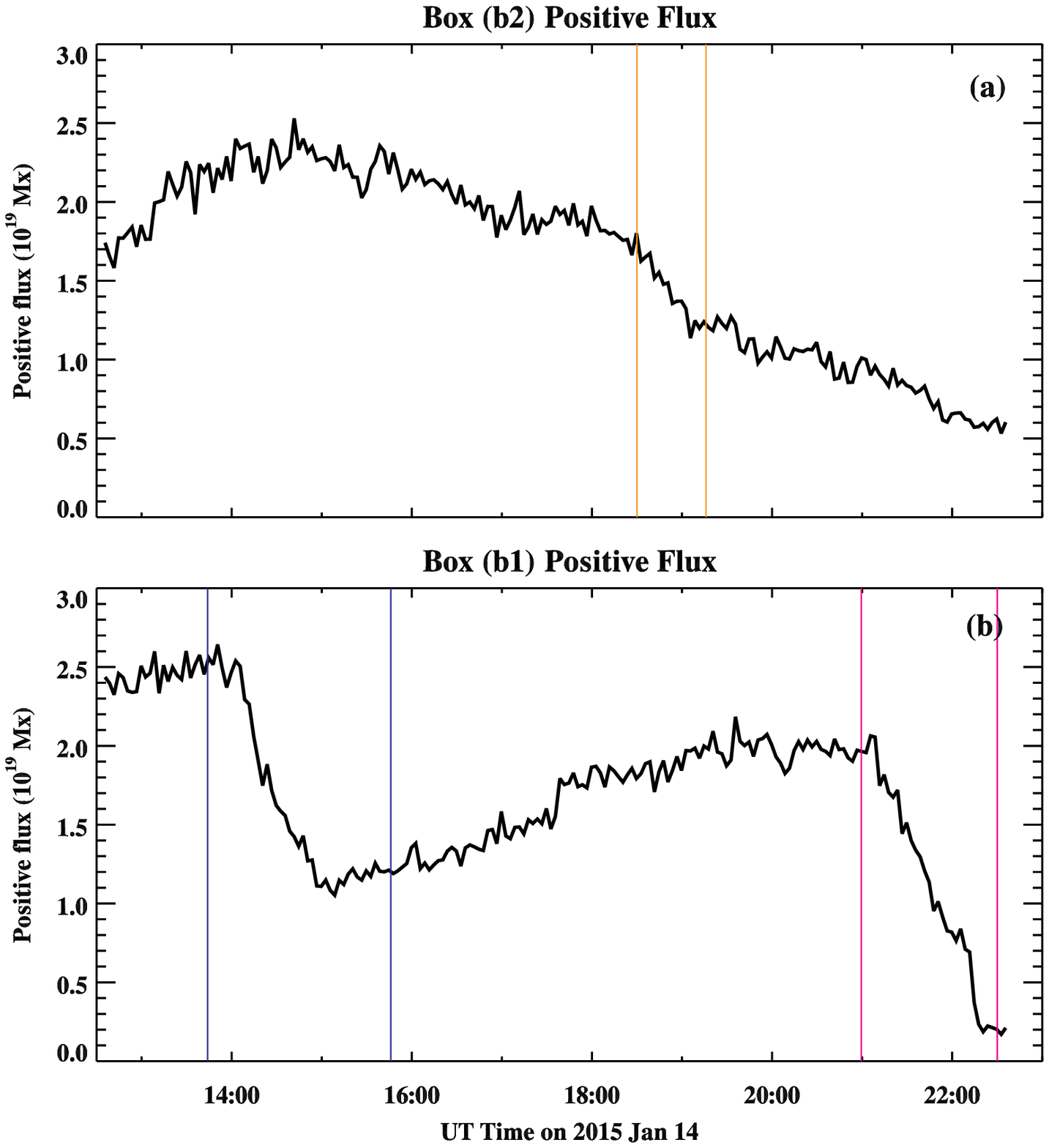}
\centerline{Figure~9}
\end{figure}
\clearpage

\end{document}